%% file: kapaev.tex
\documentclass[reqno,12pt]{amsart}
\usepackage{amsmath, euscript, amsthm}

\input psbox

\font\sc=cmcsc10 scaled \magstep1	
\font\scs=cmcsc10 	

\font\msbm=msbm10
\font\msbmsmall=msbm10 scaled 800
\font\msbmtiny=msbm10 scaled 700
\newfam\msbmfam
\textfont\msbmfam=\msbm
\scriptfont\msbmfam=\msbmsmall
\scriptscriptfont\msbmfam=\msbmtiny

\def\sgn{\mbox{\rm sgn}\,}
\def\Tr{\mbox{\rm Tr}\,}
\def\tr{\mbox{\rm\tiny Tr}\,}
\def\diag{\mbox{\rm diag}}
\def\scriptdiag{\mbox{\rm\tiny diag}\,}
\def\Re{\mbox{\rm Re}\,}
\def\Im{\mbox{\rm Im}\,}
\def\scriptRe{\mbox{\rm\tiny Re}\,}
\def\varkappa{\mbox{\msbm \char'173}}      
\def\scriptvarkappa{\mbox{\msbmsmall \char'173}}
\def\Bbb{\mathbb}
\def\cal{\mathcal}

\def\by{\times}
\def\WKB{{\scriptstyle W\!K\!B}}
\def\res{\mathop{\hbox{res}}}

\theoremstyle{plain}
\newtheorem{thm}{Theorem}[section]
\newtheorem{prop}[thm]{Proposition}
\newtheorem{cor}[thm]{Corollary}

\theoremstyle{definition}
\newtheorem{defn}{Definition}[section]

\theoremstyle{remark}
\newtheorem{rem}{Remark}[section]

\renewcommand{\theequation}{\thesection.\arabic{equation}}
\numberwithin{equation}{section}

\begin{document}
\title[Monodromy approach to the scaling limits]
{Monodromy approach to the scaling limits in the isomonodromy
systems}
\author{Andrei A. Kapaev}
\address{St Petersburg Department of Steklov Mathematical Institute
of Russian Academy of Sciences, Fontanka 27, St Petersburg, 191011,
Russia} 
\curraddr{Department of Applied Mathematics and Theoretical Physics,
University of Cambridge, Silver st., Cambridge, CB3 9EW, England}
\email{kapaev@pdmi.ras.ru, A.Kapaev@damtp.cam.ac.uk}

\begin{abstract}
The isomonodromy deformation method is applied to the scaling limits
in the linear $N\times N$ matrix equations with rational coefficients
to obtain the deformation equations for the algebraic curves which
describe the local behavior of the reduced versions for the relevant
isomonodromy deformation equations. The approach is illustrated by the
study of the algebraic curve associated to the $n$-large asymptotics
in the sequence of the bi-orthogonal polynomials with cubic
potentials. 
\end{abstract}
\maketitle

\section{Introduction}\label{sect1}

It is well known that, in certain asymptotic limits, the classical
Painlev\'e equations \cite{ince} reduce to elliptic
ones. For instance, the Painlev\'e sixth equation,
\begin{multline}\label{P6}\tag{PVI}
y_{xx}=\frac{1}{2}
\Bigl(\frac{1}{y}+\frac{1}{y-1}+\frac{1}{y-x}\Bigr)y_x^2
-\Bigl(\frac{1}{x}+\frac{1}{x-1}+\frac{1}{y-x}\Bigr)y_x+
\\
+\frac{y(y-1)(y-x)}{x^2(x-1)^2}
\Bigl(c_1+c_2\frac{x}{y^2}+c_3\frac{x-1}{(y-1)^2}
+c_4\frac{x(x-1)}{(y-x)^2}\Bigr),
\end{multline}
with the large parameters $c_j$, $j=1,\dots,4$, after the changes 
$x=t_0+\delta\tau$, $c_j=\delta^{-2}a_j+\delta^{-1}b_j$,
where $t_0,a_j,b_j=const$, turns at $\delta=0$ into an
autonomous equation which has the first integral
\begin{equation}\label{D0_P6}
D_0=\frac{t_0^2(t_0-1)^2}{2y(y-1)(y-t_0)}y_{\tau}^2
-a_1y
+a_2\frac{t_0}{y}+a_3\frac{t_0-1}{y-1}
+a_4\frac{t_0(t_0-1)}{y-t_0}.
\end{equation}
The asymptotics of the classical Painlev\'e transcendents w.r.t.\
parameters or initial data were studied in numerous works, see 
\cite{kapaev:scl1} for extended but not exhaustive bibliography. The
limit transitions of such kind are called below the {\em scaling
limits}.

The most effective to this date approach to the scaling limits in the
Painlev\'e transcendents is based on the monodromy representation for
the latter \cite{jmu, fn}. In \cite{kapaev:scl1}, the isomonodromy
deformation technique of \cite{its_nov} was adapted to the study of
the scaling limits in the equations of the isomonodromy deformations
for the $2\times2$ matrix linear first order ODEs with rational
coefficients. In particular, the results of \cite{kapaev:scl1} imply
that the modular parameters determining the limiting (hyper)elliptic
curve for the asymptotic solution of the isomonodromy deformation
equation like $D_0$ in (\ref{D0_P6}) are {\em not} arbitrary constants
but certain functions of all the {\em deformation} parameters. These
functions are uniquely determined by the system of transcendent 
{\em modulation} equations 
\begin{equation}\label{modulation_intro}
\Re\oint_{\ell}\mu(\lambda)\,d\lambda=const,
\end{equation}
where $\ell$ is an arbitrary closed path on the Riemann surface of the
relevant spectral curve.

Below, this result will be extended to the scaling limits in
the isomonodromy deformation systems for $N\times N$ matrix linear
ODEs with rational coefficients. The work is motivated by recent
developments in the theory of coupled random matrices. Indeed, while
the statistic properties of the ensembles of the single random
matrices can be given in terms of the asymptotics of the
semi-classical orthogonal polynomials, see \cite{mehta,nevai}, which
give rise to the linear first order $2\times2$ matrix ODEs with
rational coefficients \cite{fokas_its_kitaev}, the ensembles of the
coupled random matrices in the very similar way give rise to the
bi-orthogonal polynomials \cite{EMcL, BEH, Z-J} and the linear
$N\times N$ matrix ODEs \cite{BEH}.

The paper is organized as follows. In Section~\ref{basics}, we recall
the basic facts in the isomonodromy deformations of the
linear matrix equations with rational coefficients, introduce the
notion of the scaling limits in such systems and describe the WKB
approach to their asymptotic solutions. In Section~\ref{modulation},
we find the modulation equations for the non-singular asymptotic
spectral curve and prove their unique solvability. In
Section~\ref{applications}, we illustrate our approach using a
particular $3\times3$ matrix equation satisfied by the bi-orthogonal
polynomials for the cubic potentials.

\section{The isomonodromy systems with a large parameter}
\label{basics}

In this section, following \cite{fedor, wasow, jmu}, we recall the
basic notions of the theory of the linear matrix differential
equations. Consider an $N\times N$ matrix first order ODE,
\begin{equation}\label{Psi_lambda}
\frac{d\Psi}{d\lambda}=A(\lambda)\Psi,\quad
A(\lambda)=\sum_{\nu=1}^n\sum_{k=0}^{r_{\nu}}
\frac{A_{\nu,-k}}{(\lambda-a^{(\nu)})^{k+1}}-
\sum_{k=1}^{r_{\infty}}A_{\infty,-k}\lambda^{k-1}.
\end{equation}
We call equation (\ref{Psi_lambda}) {\em generic} if the eigenvalues
of $A_{\nu,-r_{\nu}}$ are distinct for $r_{\nu}\neq0$ and if they are
distinct modulo integers for $r_{\nu}=0$. Without loss of generality, 
$\lambda=\infty$ is the singular point of the highest Poincar\'e rank,
i.e. $r_{\infty}\geq r_{\nu}$, $\nu=1,\dots,n$. Assume that
(\ref{Psi_lambda}) is generic and $A_{\infty,-r_{\infty}}$
is diagonal (for the non-generic situations, see \cite{wasow}). Then,
near the singularity $a^{(\nu)}$, equation (\ref{Psi_lambda}) has the
formal solution 
\begin{equation}\label{Psi_formal}
\begin{split}
&\Psi_{\nu}(\lambda)=W^{(\nu)}\hat\Psi^{(\nu)}(\lambda)
e^{\theta^{(\nu)}(\lambda)},\quad
\nu=1,\dots,n,\infty,
\\
&\hat\Psi^{(\nu)}(\lambda)=
I+\sum_{j=1}^{\infty}\psi_j^{(\nu)}\xi^j,\quad
\theta^{(\nu)}(\lambda)=
\sum_{j=1}^{r_{\nu}}x_{-j}^{(\nu)}\frac{\xi^{-j}}{(-j)}+
x_0^{(\nu)}\ln\xi,
\end{split}
\end{equation}
where $\xi=\lambda-a^{(\nu)}$ for a finite singularity $a^{(\nu)}$
and $\xi=1/\lambda$ for infinity. The matrix coefficients
$\psi_j^{(\nu)}$ and the diagonal matrix coefficients $x_{-j}^{(\nu)}$
of the formal expansion (\ref{Psi_formal}) are determined uniquely by
the eigenvector matrix $W^{(\nu)}$ of $A_{\nu,-r_{\nu}}$.

The ratio $\tilde\Psi^{-1}(\lambda)\Psi(\lambda)$ of any two solutions 
$\Psi$ and $\tilde\Psi$ of (\ref{Psi_lambda}) does not depend on
$\lambda$. The ratios of the fundamental solutions normalized by 
(\ref{Psi_formal}) are called the {\em monodromy data}.

The set of {\em deformation parameters} is specified for generic
equation (\ref{Psi_lambda}) in \cite{jmu} (for non-generic equations
in \cite{fedor2}). These parameters include the positions 
$a^{(\nu)}$ of the singular points and the entries of the diagonal
matrices $x_{-j}^{(\nu)}$, $j\neq0$, for $\theta^{(\nu)}(\lambda)$ in
(\ref{Psi_formal}). All these quantities form together the vector $x$
of the deformation parameters. Remaining parameters $x_0^{(\nu)}$,
$\nu=1,\dots,\infty$, are called the {\em formal monodromy
exponents}. The latter satisfy the Fuchs' identity,
\begin{equation}\label{Fuchs_identity}
\sum_{\nu=1}^n\Tr x_0^{(\nu)}+\Tr x_0^{(\infty)}=0.
\end{equation}

\smallskip
\begin{rem}\label{param_reduction}
Using the linear transformations of the complex $\lambda$-plane, one
can fix the positions of two of the finite singular points or to
fix two of the entries $(x_{-j}^{(\nu)})_{kk}$. Using the scalar gauge
transformation, one can get $\Tr A(\lambda)\equiv0$ and therefore 
$\Tr x_{-j}^{(\nu)}=0$. The orbits of the above transformations are
called the {\em essential} deformation parameters. 
\end{rem}

\smallskip
Let $d$ denote the exterior differentiation w.r.t.\ entries of $x$. In
accord with \cite{jmu}, the monodromy data of the generic equation
(\ref{Psi_lambda}) do not depend on $x$ if and only if there exist
1-forms $\Omega$ and $\Theta^{(\nu)}$ such that the fundamental
solutions above additionally satisfy equations 
\begin{equation}\label{dPsi}
d\Psi=\Omega\Psi,\quad
dW^{(\nu)}=\Theta^{(\nu)}W^{(\nu)},\quad
\nu=1,\dots,n,\infty.
\end{equation}
Then the compatibility condition of (\ref{Psi_lambda}),
(\ref{dPsi}),
\begin{equation}\label{Frobenius}
dA=\frac{\partial\Omega}{\partial\lambda}+[\Omega,A],\quad
d\Omega=\Omega\wedge\Omega,
\end{equation}
is the completely integrable differential system whose fixed
singularities are the planes $a^{(\nu)}=a^{(\rho)}$, $\nu\neq\rho$, 
$(x_j^{(\nu)})_{ii}=\infty$, $(x_{-r_{\nu}}^{(\nu)})_{ii}=0$,
$(x_{-r_{\nu}}^{(\nu)})_{ii}=(x_{-r_{\nu}}^{(\nu)})_{jj}$ if
$i\neq j$, $r_{\nu}\neq0$. The generic $2\times2$ system
(\ref{Frobenius}) admitting the only one deformation parameter is
equivalent to one of the classical Painlev\'e equations.

\subsection{Scaling limits in the isomonodromy systems}

Let $A(\lambda)$ depend on an additional parameter $\delta$,
\begin{equation}\label{scaling_change}
\begin{split}
&a^{(\nu)}=\delta^{\scriptvarkappa}(b^{(\nu)}+\delta c^{(\nu)}),\quad
\varkappa=const,\quad
\nu=1,\dots,n,
\\
&A_{\nu,-k}=\delta^{k\scriptvarkappa-1}B_{\nu,-k},\quad
A_{\infty,-k}=\delta^{-k\scriptvarkappa-1}B_{\infty,-k},\quad
k=0,\dots,r_{\nu},
\end{split}
\end{equation}
where $B_{\nu,-k}$, $k=0,\dots,r_{\nu}$,
$\nu=1,\dots,n,\infty$, are ascending Erdelyi series in
$\delta$ such that, for generic equation (\ref{Psi_lambda}),
\begin{equation}\label{x_change}
x_{-k}^{(\nu)}=\delta^{k\scriptvarkappa-1}
(t_{-k}^{(\nu)}+\delta\tau_{-k}^{(\nu)}),\quad
x_{-k}^{(\infty)}=
\delta^{-k\scriptvarkappa-1}
(t_{-k}^{(\infty)}+\delta\tau_{-k}^{(\infty)}).
\end{equation}
The constants $b^{(\nu)}$ and the entries of $t_{-k}^{(\nu)}$,
$k\neq0$, form the vector $t$ for the center of the asymptotic domain
in the parameter space as $\delta\to+0$. Inequalities
$|c^{(\nu)}|,\|\tau_{-k}^{(\nu)}\|<const$, or simply
$\|\tau\|<const$, yield the range of the ``local'' deformations. The
entries of $t$ are usually called the ``slow'' variables, while the
entries of $\tau$ are called the ``quick'' or ``fast'' variables. The
entries of the vectors $t_0^{(\nu)}$ and $\tau_0^{(\nu)}$ for the
formal monodromy exponents form the vectors $\alpha$ and $\beta$,
respectively.

Substituting (\ref{scaling_change}) and
$\lambda=\delta^{\scriptvarkappa}\zeta$ into (\ref{Psi_lambda}),
(\ref{dPsi}), we find 
\begin{align}\label{Psi_zeta}
&&&\frac{d\Psi}{d\zeta}=\delta^{-1}B(\zeta)\Psi,\quad
d\Psi=\omega\Psi,
\\
&&&B(\zeta)=\sum_{\nu=1}^n\sum_{k=0}^{r_{\nu}}
\frac{B_{\nu,-k}}{(\zeta-b^{(\nu)}-\delta c^{(\nu)})^{k+1}}-
\sum_{k=1}^{r_{\infty}}B_{\infty,-k}\zeta^{k-1},\notag
\end{align}
whose compatibility reads
\begin{equation}\label{scaled_Frobenius}
dB=[\omega,B]+\delta\frac{\partial\omega}{\partial\zeta},\quad
d\omega=\omega\wedge\omega.
\end{equation}

\smallskip
\begin{rem}\label{Remark1} If $r_{\infty}\neq0$ and 
$(x_{-r_{\infty}}^{(\infty)})_{11}$ does not depend on $\delta$ then
$\varkappa=-1/r_{\infty}$. If all the singularities are Fuchsian,
i.e.\ $r_{\nu}=0$, $\nu=1,\dots,n,\infty$, one may put for simplicity
$\varkappa=0$.
\end{rem}

\smallskip
\begin{rem}\label{Remark2} Following \cite{jm2, fokas_its_kitaev}, it
is possible to construct a Schlesinger transformation chain yielding
$x_0^{(\nu)}=\rho^{(\nu)}+l^{(\nu)}$ with constant
$\rho^{(\nu)}\in{\Bbb C}^N$ and $l^{(\nu)}\in{\Bbb Z}^N$. Taking
$\|l^{(\nu)}\|$ and $\delta^{-1}$ large and comparable,
$\lim_{\delta\to+0}l^{(\nu)}\,\delta=t_0^{(\nu)}=const$, one writes
$x_0^{(\nu)}=\delta^{-1}t_0^{(\nu)}+\tau_0^{(\nu)}$ where
$\tau_0^{(\nu)}\in{\Bbb C}^N$ is bounded as $\delta\to+0$, and 
\begin{equation}\label{reality_t0}
t_0^{(\nu)}\in{\Bbb R}^N. 
\end{equation}
This allows us to interpret $\Re(t_0^{(\nu)})_{jj}$ as the {\em
discrete} deformation parameters with the step ${\cal O}(\delta)$, see
below.
\end{rem}

\begin{rem}\label{scaling_freedom}
The scaling parameter $\delta$ is defined up to a positive factor
which gives rise to a scaling freedom in the set of the ``slow''
variables. Below, we assume that this scaling freedom is eliminated by
a normalization of one of the non-trivial ``slow'' deformation
parameters.
\end{rem}

\subsection{Complex WKB method}\label{wkb_method}

Here we recall the idea of the complex WKB method following in
principal \cite{fedor, wasow}. Consider (\ref{Psi_zeta}) as
$\delta\to0$ assuming that the coefficients of $B(\zeta)$ remain
bounded. Let $T$ and $\Lambda_0$ be the eigenvector and eigenvalue
matrices for $B(\zeta)$, 
\begin{equation}\label{recurrence_for_T0_L0}
T^{-1}BT=\Lambda_0=\diag(\mu_1,\dots,\mu_N).
\end{equation}
Let $\mu_j\neq\mu_k$, $j\neq k$. Then the formal expression
\begin{equation}\label{w22f}
\Psi(\zeta)=T\sum_{n=0}^{\infty}\delta^nT_n
\exp\Bigl\{\delta^{-1}\int_{\zeta_0}^{\zeta}
\sum_{m=0}^{\infty}\delta^m\Lambda_m(\xi)\,d\xi\Bigr\},\quad
T_0=I,
\end{equation}
satisfies (\ref{Psi_zeta}) provided the diagonal matrices $\Lambda_n$
and the off-diagonal matrices $T_n$, $n\geq1$, solve the recursion
\begin{align}
&&&[\Lambda_0,T_1]-\Lambda_1=
T^{-1}\frac{dT}{d\lambda},\notag
\\\label{recurrence_for_Tn_Ln}
&&&[\Lambda_0,T_n]-\Lambda_n=
\sum_{m=1}^{n-1}T_{n-m}\Lambda_m
+T^{-1}\frac{d(TT_{n-1})}{d\lambda},\quad
n\geq2.
\end{align}

Let $\Gamma$ be the Riemann surface of the algebraic curve
\begin{equation}\label{spectral_curve}
F(\zeta,\mu):=\det(B(\zeta)-\mu I)=0.
\end{equation}
The branch points of (\ref{spectral_curve}) are called the 
{\em turning points} for (\ref{Psi_zeta}). Let ${\cal L}_0$ be an
{\em open\/} simply connected domain which is the complex $\zeta$-plane
punctured at the singularities of $B(\zeta)$, at the turning points
and cut along the segments connecting all the singularities and
turning points. Let ${\cal L}\subset{\cal L}_0$ be a {\em closed\/}
simply connected domain. By construction, $B(\zeta)$ is holomorphic in
${\cal L}$, and equation (\ref{Psi_zeta}) has no turning point in
${\cal L}$. Thus all the roots $\mu_j(\lambda)$ of the characteristic
equation (\ref{spectral_curve}) are distinct from each other and are
holomorphic in ${\cal L}$. Therefore there exists a holomorphic
non-special in  ${\cal L}$ matrix $T(\zeta)$ which diagonalizes the
matrix $B(\zeta)$, $T^{-1}BT=\Lambda_0=\diag(\mu_1,\dots,\mu_N)$. The
matrices $T$, $T^{-1}$, $\Lambda_n$, $T_n$, are holomorphic and
bounded in ${\cal L}$ \cite{fedor}.

Consider the reduced gauge matrix
\begin{equation}\label{w28}
{\cal T}^{(m)}=T\sum_{n=0}^{m}\delta^nT_n,\quad
T_0=I,
\end{equation}
where $T_n$ are defined by 
(\ref{recurrence_for_T0_L0})--(\ref{recurrence_for_Tn_Ln}), and 
$\zeta\in{\cal L}$. The matrix function 
$\Phi^{(m)}=({\cal T}^{(m)})^{-1}\Psi$ solves the ``almost diagonal''
equation 
\begin{equation}\label{Phi_eq}
\Phi_{\zeta}^{(m)}=\delta^{-1}B^{(m)}\Phi^{(m)},\quad
B^{(m)}(\zeta)=\sum_{n=0}^{m}\delta^n\Lambda_n+\delta^{m+1}R^{(m)},
\end{equation}
where $R^{(m)}(\zeta)$ is holomorphic and bounded for 
$\zeta\in{\cal L}$ provided $\delta$ is small enough. Define the WKB
approximation to (\ref{Phi_eq}),
\begin{equation}\label{w212}
\Phi_{\WKB}^{(m)}(\zeta)=e^{\theta(\zeta_0,\zeta)},\quad
\theta(\zeta_0,\zeta)=\delta^{-1}\int_{\zeta_0}^{\zeta}
\Lambda^{(m)}\,d\zeta,\quad
\Lambda^{(m)}=\sum_{n=0}^{m}\delta^n\Lambda_n,
\end{equation}
and introduce the correction function $\chi^{(m)}$,
\begin{equation}\label{chi^(m)_def}
\Phi^{(m)}(\zeta)=\chi^{(m)}(\zeta)\Phi_{\WKB}^{(m)}(\zeta),
\end{equation}
and notations
\begin{equation}\label{notations}
\begin{split}
&\mu_{ij}=\mu_i-\mu_j,\quad
\theta_{ij}(\xi,\zeta)=
\bigl(\theta(\xi,\zeta)\bigr)_{ii}
-\bigl(\theta(\xi,\zeta)\bigr)_{jj},
\\
&\hat\theta_{ij}(\xi,\zeta)=
\theta_{ij}(\xi,\zeta)
-\delta^{-1}\int_{\xi}^{\zeta}\mu_{ij}(s)\,ds.
\end{split}
\end{equation}

The contour $\gamma_{ij}(\zeta)\subset{\cal L}$ connecting the finite
or infinite point $\zeta_{ij}$ with $\zeta$ is called the 
{\em $(i,j)$-canonical\/} path if 
\begin{equation}\label{can_path_def}
\Re\int_{\xi}^{\zeta}\mu_{ij}(s)\,ds\leq0\quad
\forall\xi\in\gamma_{ij}(\zeta).
\end{equation}
A closed simply connected domain ${\cal C}_{ij}\subset{\cal L}$ is
called {\em $(i,j)$-canonical} if there exists such a point 
$\zeta_{ij}\in{\cal C}_{ij}$ that the contour 
$\gamma_{ij}(\zeta)\subset{\cal C}_{ij}$ connecting 
$\zeta_{ij}$ with any given point $\zeta\in{\cal C}_{ij}$ is
homotopy equivalent to a canonical path. A closed simply connected
domain ${\cal C}$ is called {\em canonical\/} if it is
$(i,j)$-canonical $\forall i,j=1,\dots,N$, $i\neq j$.

Using the arguments of \cite{fedor}, we thus obtain the following
\begin{thm}\label{Theorem1}
Let ${\cal C}\subset{\cal L}$ be a canonical domain. If
$$
\int_{\gamma_{ij}(\zeta)}
e^{\scriptRe\hat\theta_{ij}(\xi,\zeta)}
\bigl|\bigl(R^{(m)}(\xi)\bigr)_{ij}\bigr|
\cdot|d\xi|<\infty,\quad
\forall i,j=1,\dots,N,\quad
\forall\zeta\in{\cal C},$$ 
then there exist such positive constants $C$ and $\delta_0$ that
\begin{equation}\label{WKB_estim}
\|\chi^{(m-1)}(\zeta)-I\|\leq C|\delta|^{m}\quad
\forall\zeta\in{\cal C},\quad
\forall\delta\in(0,\delta_0].
\end{equation}
\end{thm}

In particular, $\chi^{(0)}(\zeta)=I+{\cal O}(\delta)$.

To construct a canonical domain containing a given point
$\zeta_0\in{\cal L}$, consider a pair of $i,j\in\{1,\dots,N\}$,
$i\neq j$, and introduce the segment $\ell_{ij}\subset{\cal L}$ of the 
{\em $(i,j)$-anti-Stokes level curve-line} passing through $\zeta_0$:
\begin{equation}\label{aStokes_ll_def}
\ell_{ij}=\bigl\{\zeta\in{\cal L}\colon\
\Im\int_{\zeta_0}^{\zeta}\mu_{ij}(s)\,ds=0\bigr\}.
\end{equation}
Choose two points $\zeta_{ij},\zeta_{ji}\in\ell_{ij}$ in such a way
that:
a)~$\zeta_0\in[\zeta_{ij},\zeta_{ji}]$;  
b)~for any $\zeta\in\ell_{ij}$ separating $\zeta_{ij}$ from $\zeta_{ji}$,
the curve-line segment $[\zeta_{ij},\zeta]\subset\ell_{ij}$ is the 
$(i,j)$-canonical path, while $[\zeta_{ji},\zeta]\subset\ell_{ij}$
is the $(j,i)$-canonical path. Introduce the segment
$\ell_{ij}^*\subset{\cal L}$ of the {\em $(i,j)$-Stokes level
curve-line} passing through the point 
$\zeta^*\in[\zeta_{ij},\zeta_{ji}]\subset\ell_{ij}$,
\begin{equation}\label{Stokes_ll_def}
\ell_{ij}^*=\bigl\{\zeta\in{\cal L}\colon\
\Re\int_{\zeta^*}^{\zeta}\mu_{ij}(s)\,ds=0\bigr\}.
\end{equation}
By construction, the union ${\cal C}_{ij}$ of all the curve-line
segments $\ell_{ij}^*$,
\begin{equation}\label{C_ij_def}
{\cal C}_{ij}=\bigcup_{\zeta^*\in[\zeta_{ij},\zeta_{ji}]}\ell_{ij}^*,
\end{equation}
is the $(i,j)$- and $(j,i)$-canonical domain. The boundary of the
constructed $(i,j)$-canonical domain ${\cal C}_{ij}$ is formed by the
$(i,j)$-Stokes level curve-lines passing through the points
$\zeta_{ij}$ and $\zeta_{ji}$ and partially by the boundary of 
${\cal L}$. It is worth to note that, near the irregular singularities
of (\ref{Psi_zeta}), the $(i,j)$-canonical domain ${\cal C}_{ij}$ can
be extended beyond the boundary of ${\cal L}$ to fill out certain
sector in the complex $\zeta$-plane called the {\em $(i,j)$-Stokes
sector}.

The canonical domain ${\cal C}\ni\zeta_0$ of validity of
Theorem~\ref{Theorem1} is the intersection of the above
$(i,j)$-canonical domains ${\cal C}_{ij}$,
\begin{equation}\label{C_def}
{\cal C}=\bigcap_{\stackrel{i,j\in\{1,\dots,N\}}{i\neq j}}{\cal C}_{ij}.
\end{equation}

The above construction implies that any closed simply connected domain
${\cal L}\subset{\cal L}_0$ can be covered by a {\em finite} number of
the {\em overlapping} canonical domains ${\cal C}^{(k)}$,
$k=1,\dots,K$, since the opposite assumption can be easily brought to
a contradiction.

\section{Modulation of the spectral curve}\label{modulation}

Solutions of the Lax equation $dB=[\omega,B]$ are routinely
interpreted as the approximate solutions for (\ref{scaled_Frobenius})
as $\delta\to0$ \cite{garnier2, FN2}. Supplemented by the
eigenvalue problem (\ref{recurrence_for_T0_L0}), the Lax equation
constitutes the basis for the algebro-geometric integration of the
``soliton'' equations. However, in the theory of the ``soliton'' PDEs,
the spectral curve (\ref{spectral_curve}) is determined by the initial
data, while in the isomonodromy deformation context it is determined
by the original $\lambda$-equation (\ref{Psi_lambda}). Moreover
the spectral curve for the typical ``soliton'' equation is an exact
integral of motion while, in the isomonodromy case, the curve varies, 
$$
d(\ln\det(B-\mu I))=\delta\,\Tr(
\frac{\partial\omega}{\partial\zeta}(B-\mu I)^{-1})\not\equiv0.
$$
In what follows, we precisely describe the dependence of the algebraic
curve (\ref{spectral_curve}) on the ``slow'' variables at
$\delta=0$. Below, the subscript $_{as}$ denotes the relevant object
at $\delta=0$. 

We call the curve (\ref{spectral_curve}) {\em singular} if its
topological properties for all small enough $\delta\neq0$ differ from
those at $\delta=0$. Given a parameterization of the curve, we define
the {\em discriminant set} ${\mathfrak S}$ in the total parameter space
${\mathfrak P}$ as the set determining the singular
curve. Also, let ${\mathfrak F}$ be the union of the hyper-planes
$b^{(\nu)}=b^{(\rho)}$, $\nu\neq\rho$, 
$(t_{-r_{\nu}}^{(\nu)})_{ii}=(t_{-r_{\nu}}^{(\nu)})_{jj}$, $i\neq j$,
$(t_{-r_{\nu}}^{(\nu)})_{kk}=0$ corresponding to the fixed
singularities of (\ref{scaled_Frobenius}) at $\delta=0$. Below, we
always assume that our deformation parameters are apart from the fixed
singularities ${\mathfrak F}$.

The differential $\mu_{as}(\zeta)\,d\zeta$ as well as its derivatives
w.r.t.\ $b^{(\nu)}$ and entries of $t_{-k}^{(\nu)}$,
$k=0,\dots,r_{\nu}$, $\nu=1,\dots,n,\infty$, are meromorphic on the
Riemann surface $\Gamma_{as}$ of the curve. All the parameters
$b^{(\nu)},t_{-k}^{(\nu)}$ together completely determine the singular
part of $\mu_{as}(\zeta)\,d\zeta$.
\begin{defn}
The parameter $D_j$ is called {\em modular} iff the differential 
$\frac{\partial\textstyle}{\partial D_j}\mu_{as}(\zeta)\,d\zeta$ is
holomorphic on the Riemann surface $\Gamma_{as}$.
\end{defn}
Thus ${\mathfrak P}={\mathfrak T}\otimes{\mathfrak D}$, where
${\mathfrak T}$ is the subspace of the deformation parameters 
${\bf t}=(t,\Re\alpha)$ (see Remark~\ref{Remark2} and constraint
(\ref{Fuchs_identity})) while ${\mathfrak D}$ is the subspace of the
remaining parameters ${\bf D}=(D,\Im\alpha)$.

\begin{thm}\label{modulation_gen}
Let $({\bf t}_0,{\bf D}_0)\in{\mathfrak P}\setminus{\mathfrak S}$.  
Then there exists an open neighborhood\/ 
${\cal U}\subset{\mathfrak T}\setminus{\mathfrak F}$ of\/ ${\bf t}_0$
such that, for any closed path $\ell$ on the Riemann surface
$\Gamma_{as}$ punctured over the points $b^{(\nu)}$,
$\nu=1,\dots,n,\infty$, 
\begin{equation}\label{modulation_equation}
J_{\ell}({\bf t},{\bf D}):=\Re\oint_{\ell}\mu_{as}(\zeta)\,d\zeta=
h_{\ell}\quad
\forall{\bf t}\in{\cal U},
\end{equation}
where $h_{\ell}=J_{\ell}({\bf t}_0,{\bf D}_0)=const$.
\end{thm}

\proof
Let $\ell_{\zeta}=\pi(\ell)$ be a projection of the closed path
$\ell\subset\Gamma$ with the base point $(\zeta_0,\mu_0)$ on the
punctured complex $\zeta$-plane. Let the integer $m_0$ be chosen in
such a way that the lift $\hat\ell$ of $m_0\ell_{\zeta}$ on $\Gamma$
be closed for all branches of $\mu(\zeta_0)$. Consider the analytic
continuation of the WKB approximation (\ref{w22f}) along
$\hat\ell$. The projection $\pi(\hat\ell)=m_0\ell_{\zeta}$ is covered
by a finite number of the overlapping canonical domains ${\cal C}_k$,
$k=1,\dots,s$, ${\cal C}_{s+1}={\cal C}_1$, in each of which
(\ref{w22f}) approximates uniformly in $\zeta$ an exact solution
$\Psi_k(\zeta)$ of (\ref{Psi_zeta}). Because
$\Psi_{k+1}(\zeta)=\Psi_k(\zeta)G_k$ where $G_k$ is independent from
both $\zeta$ and ${\bf t}\in{\mathfrak T}\setminus{\mathfrak F}$, we
obtain 
$$
{\cal M}_{\hat\ell}(\Psi_{s+1}(\zeta))=
\Psi_1(\zeta)M_{\hat\ell}G_1\cdots G_s,
$$ 
where ${\cal M}_{\hat\ell}$ is the operator of analytic
continuation along $m_0\ell_{\zeta}$, and $M_{\hat\ell}$ is the
monodromy matrix for $\Psi_1(\zeta)$ along $m_0\ell_{\zeta}$. 
Using for $\Psi_1(\zeta)$ and $\Psi_{s+1}(\zeta)$ our WKB
approximation, we find 
\begin{equation}\label{oint_of_Lambda}
\exp\Bigl\{\delta^{-1}\oint_{\hat\ell}\Lambda(\zeta)\,d\zeta\Bigr\}=
(I+{\cal O}(\delta))G(\delta).
\end{equation}
Since the curve is non-singular, the r.h.s.\ of (\ref{oint_of_Lambda})
preserves while ${\bf t}$ remains in a neighborhood of ${\bf t}_0$.
Equating the leading orders of the l.h.s.\ for (\ref{oint_of_Lambda})
at ${\bf t}_0$ and nearby points ${\bf t}$, we arrive at 
(\ref{modulation_equation}). 
\endproof

Theorem~\ref{modulation_gen} immediately provides us with the
following assertion:

\begin{cor}\label{corollary1}
Let\/ ${\cal U}\subset{\mathfrak T}\setminus{\mathfrak F}$ be an open
domain and let (\ref{modulation_equation}) holds true. If the curve
$F(\zeta,\mu)=0$ remains non-singular at the boundary point
${\bf t}_1\in\partial{\cal U}$, then there exists an open domain\/
${\cal W}\subset{\mathfrak T}\setminus{\mathfrak F}$ such that 
${\cal U}\subset{\cal W}$, ${\bf t}_1\in{\cal W}$, and
(\ref{modulation_equation}) is valid $\forall{\bf t}\in{\cal W}$.
\end{cor}

For the subsequent discussion, the following assertion is useful:
\begin{prop}\label{J_continuity}
For any closed path $\ell$ on $\Gamma_{as}$ punctured over 
$b^{(\nu)}$, $\nu=1,\dots,n,\infty$, the integral 
$J_{\ell}({\bf t},{\bf D})$ is continuous in $({\bf t},{\bf D})$
outside the fixed singularities ${\mathfrak F}$ of
(\ref{scaled_Frobenius}) and is differentiable in $({\bf t},{\bf D})$ 
outside the discriminant set ${\mathfrak S}$. 
\end{prop}

\begin{proof}
Since the contour $\ell$ is finite, the continuity of 
$J_{\ell}({\bf t},{\bf D})$ follows from the continuity of
$\mu_{as}(\zeta)$. If the point $({\bf t}_0,{\bf D}_0)$ is located
apart from the discriminant set ${\mathfrak S}$, then there exists an
open neighborhood ${\cal V}$ of $({\bf t}_0,{\bf D}_0)$ such that the
spectral curve (\ref{spectral_curve}) does not 
degenerate $\forall({\bf t},{\bf D})\in{\cal V}$. Then the 
differentiability of $J_{\ell}({\bf t},{\bf D})$ at 
$({\bf t}_0,{\bf D}_0)$ follows from the continuous differentiability
of $\mu_{as}(\zeta)$. 
\end{proof}

Theorem~\ref{modulation_gen} and Proposition~\ref{J_continuity} imply
\begin{cor}\label{corollary2}
Let\/ 
${\cal U},\hat{\cal U}\subset{\mathfrak T}\setminus{\mathfrak F}$ be 
adjacent open domains and let\/
$(\partial{\cal U}\cap\partial\hat{\cal U})\subset
{\mathfrak T}\setminus{\mathfrak F}$ be not empty. 
If $J_{\ell}({\bf t},{\bf D})=h_{\ell}\>$\ \ 
$\forall{\bf t}\in{\cal U}$ and
$J_{\ell}({\bf t},{\bf D})=\hat h_{\ell}\>$\ \ 
$\forall{\bf t}\in\hat{\cal U}$, then $h_{\ell}=\hat h_{\ell}$.
\end{cor}

In accord with Corollaries~\ref{corollary1} and~\ref{corollary2}, 
if the spectral curve (\ref{spectral_curve}) is non-singular at the
initial point $({\bf t}_0,{\bf D}_0)$, then the modulation equation
(\ref{modulation_equation}) is valid in a domain ${\cal U}$ bounded by
the points where the spectral curve becomes singular. Applicability of
(\ref{modulation_equation}) to a particular solution of
(\ref{scaled_Frobenius}) beyond this boundary depends on some subtle
details in the initial data or, equivalently, in the relevant
monodromy data of the isomonodromy system (\ref{Psi_zeta}), see
\cite{kapaev:scl1} and Section~\ref{applications} below.

Let us discuss now the existence of the function ${\bf D}({\bf t})$
such that $J_{\ell}({\bf t},{\bf D}({\bf t}))\equiv const$.
Varying the contour $\ell$ in (\ref{modulation_equation}), we obtain
the system of equations $J_{\ell_j}=h_j$, where the set of contours
$\ell_j$ form a homology basis of the Riemann surface of $\Gamma_{as}$
punctured over $b^{(\nu)}$. For instance, taking for $\ell$ a small
circle $c_{\nu}$ around $\bigl(b^{(\nu)}\!,\,\mu_j(b^{(\nu)})\bigr)$,
we find
\begin{equation}\label{Im_t0=c}
\Im(t_0^{(\nu)})_{jj}=-\tfrac{1}{2\pi}h_{c_{\nu}}^{(j)}=const,
\end{equation}
which contains (\ref{reality_t0}) as the particular case
$h_{c_{\nu}}^{(j)}=0$.

To discuss (\ref{modulation_equation}) further, it is convenient to
impose the conditions (\ref{Im_t0=c}) and to remove $(N-1)(n+1)$ small
circles from the ``sufficient" set of contours. The remaining cycles
$\ell_j$, $j=1,\dots,2g$, form a homology basis of $\Gamma_{as}$. 
Also, using (\ref{Im_t0=c}), we exclude the constant parameters 
$\Im t_0^{(\nu)}$ from the set of unknowns and assume below that the
space ${\mathfrak D}$ is $g$-dimensional complex space of the modular
parameters. 
\begin{thm}\label{D(t)}
Let $({\bf t}_0,D_0)\in{\mathfrak P}\setminus{\mathfrak S}$. Then, in
an open neighborhood\/ 
${\cal U}\subset{\mathfrak T}\setminus{\mathfrak F}$ of the point
${\bf t}_0$, the system (\ref{modulation_equation}), where $\ell$ runs
over the homology basis $\{\ell_j\}_1^{2g}$ of\/ $\Gamma_{as}$,
determines the unique differentiable in the real sense complex vector
function  $D({\bf t},\bar{\bf t})$ such that 
$D({\bf t}_0,\bar{\bf t}_0)=D_0$.
\end{thm}
\begin{proof}
Theorem~\ref{modulation_gen} and Proposition~\ref{J_continuity} imply
that $J_{\ell_j}({\bf t},D)$ are the first integrals of the completely
integrable Pfaffian system $dJ=0$, 
\begin{equation}\label{Pfaffian}
\omega\begin{pmatrix}
dD\\ d\bar D\end{pmatrix}=
-\Omega\begin{pmatrix}
d{\bf t}\\ d\bar{\bf t}
\end{pmatrix},
\end{equation}
where $\omega$ and $\Omega$ are the matrices of the partial
derivatives of $J_{\ell_j}({\bf t},D)$ w.r.t.\ the entries of the
vectors $D$, $\bar D$ and ${\bf t},\bar{\bf t}$, respectively. Here,
the bar means the complex conjugation. Let constant $c_1$ be the
determinant of the transformation of the natural basis 
$\{\frac{\partial}{\partial D_j}\mu_{as}\,d\zeta\}_{j=1}^g$
into the basis of the normalized holomorphic differentials, and let
$\hat B$ be the matrix of the $B$-periods of the normalized
holomorphic differentials. Since
$\omega=
\bigl(\begin{smallmatrix}A&\bar A\\B&\bar B\end{smallmatrix}\bigr)$
is the matrix of $A$- and $B$-periods of the holomorphic
differentials and their complex conjugate, 
$\det\omega=(-2i)^g|c_1|^2\det\bigl(\Im\hat B\bigr)\neq0$. Thus the
matrix $\omega$ is invertible until $F(\zeta,\mu)=0$ remains
non-singular, and therefore the integral manifold for (\ref{Pfaffian})
is well parameterized by the deformation parameters 
$(t,\bar t,\Re\alpha)$. 
\end{proof}

\begin{rem}\label{global_D} 
If $h_{\ell_j}=J_{\ell_j}({\bf t}_0,D_0)\neq0$ then the cycle $\ell_j$
can {\em not\/} collapse, and the encircled by $\ell_j$ branch points
can {\em not\/} coalesce. Thus, along the integral manifold for
(\ref{Pfaffian}), the spectral curve remains non-singular provided
$h_{\ell_j}\neq0$ $\forall j=1,\dots,2g$. This observation ensures the
applicability of (\ref{modulation_equation}), (\ref{Pfaffian}) and the
existence of $D({\bf t},\bar{\bf t})$ in {\em any} connected domain
${\cal U}\subset{\mathfrak T}\setminus{\mathfrak F}$ containing the 
initial point ${\bf t}_0$. 
\end{rem}

Let $J_{\ell}({\bf t},{\bf D})$ in (\ref{modulation_equation}) vanish
for {\em all\/} closed paths,
\begin{equation}\label{Boutroux} 
\Re\oint_{\ell}\mu_{as}(\zeta)\,d\zeta=0,\quad
\forall\ell\subset\Gamma_{as}.
\end{equation}
This system does make sense regardless the choice of the initial point
since there is no need to fix a homology basis. Traditionally, it is
called the {\em Boutroux system}. We recall that (\ref{Boutroux})
may be not applicable to a particular solution of
(\ref{scaled_Frobenius}) in certain sectors of 
${\mathfrak P}\setminus{\mathfrak S}$ in spite of the Boutroux system
itself does make sense in the whole parameter space ${\mathfrak P}$
(using Proposition~\ref{J_continuity}, the system (\ref{Boutroux}) is 
interpreted at the points of the discriminant set ${\mathfrak S}$ as a
continuation from ${\mathfrak P}\setminus{\mathfrak S}$).
\begin{rem}\label{discriminant_co_dim}
As the integral manifold for (\ref{Boutroux}) meets the discriminant
set ${\mathfrak S}$, at least one of the cycles $\ell_j$ collapses,
and the corresponding {\em real} equation in (\ref{Boutroux})
becomes trivial as being replaced by the {\em complex} condition of
coalescence of two branch points. Thus, generically, the intersection
of the integral manifold for (\ref{Boutroux}) with ${\mathfrak S}$ has
codim$_{\Bbb R}=1$ in the space of the deformation parameters.
\end{rem}

\begin{thm}\label{Boutroux_D}
There exists the unique solution $D({\bf t},\bar{\bf t})$ of the
Boutroux system (\ref{Boutroux}).
\end{thm}
\begin{proof}
Here, we give the sketch proof.

Uniqueness. Given ${\bf t}$, two solutions $D$ and $D'$ determine two
differentials $\mu_{as}\,d\zeta$ and $\mu_{as}'\,d\zeta$ meromorphic
on the respective Riemann surfaces $\Gamma_{as}$ and
$\Gamma_{as}'$. The difference $\phi=(\mu_{as}-\mu_{as}')\,d\zeta$ is
holomorphic on the covering Riemann surface ${\cal G}_{as}$, and
$\Re\oint_{\ell}\phi=0$ for all closed paths 
$\ell\subset{\cal G}_{as}$. However, there is no differential $\phi$
with such properties.

Existence.
Choose a point $({\bf t}_0,D_0)\in{\mathfrak P}\setminus{\mathfrak S}$
in such a way that $h_j^{(0)}=J_{\ell_j}({\bf t}_0,D_0)\neq0$ for all
contours $\ell_j$ of a homology basis $\{\ell_j\}_1^{2g}$. Consider
the extension of (\ref{Pfaffian}) where $2g$ real parameters
$h=(h_1,\dots,h_{2g})^T$ are added to the set of the independent
variables,
$$
\omega\begin{pmatrix}
dD\\ d\bar D\end{pmatrix}=
-\Omega\begin{pmatrix}
d{\bf t}\\ d\bar{\bf t}\end{pmatrix}+dh.
$$
Applying the arguments used in the proof of Theorem~\ref{D(t)} and
taking into account Remark~\ref{global_D}, we establish the 
existence of the function $D({\bf t},\bar{\bf t},h)$
$\forall{\bf t}\in{\cal U}\subset{\mathfrak T}\setminus{\mathfrak F}$
and $\forall h\colon h_j\,\sgn(h_j^{(0)})>0,\ j=1,\dots,2g$. The
assumption that $D({\bf t},\bar{\bf t},h)$ is unbounded as $h\to0$
leads to a contradiction. From a bounded sequence
$D^{(k)}=D({\bf t},\bar{\bf t},h^{(k)})$,
$h^{(k)}\to0$, we extract a convergent subsequence,
$\lim_{m\to\infty}D^{(k_m)}=D^*$. Then the continuity of the integrals
$J_{\ell_j}({\bf t},D)$ w.r.t.\ $D$ yields 
$J_{\ell_j}({\bf t},D^*)=0$. 
\end{proof}

\begin{rem}\label{Boutroux_importance}
Assuming that the monodromy data of the system (\ref{Psi_zeta}) are
generic and do not depend on the scaling parameter $\delta^{-1}$, or
this dependence is weak enough, it is possible to prove that the
relevant spectral curve satisfies the Boutroux system
(\ref{Boutroux}). Here, we do not prove this assertion (look for more
details and for the proof of this statement in the case $N=2$ in
\cite{kapaev:scl1}).
\end{rem}

\section{Modulation equations and the asymptotics of the bi-orthogonal
polynomials}\label{applications}

Modulation equations (\ref{modulation_equation}), (\ref{Boutroux}) for
the classical Painlev\'e equations were studied in 
\cite{kapaev:scls,kapaev:scl1}. Here we note that the linear ODEs
associated to the classical Painlev\'e equations 
\cite{jmu,fn,its_nov} as well as the similar equations associated to
the semi-classical orthogonal polynomials have matrix dimension
$2\times2$ \cite{fokas_its_kitaev}, thus all the relevant spectral
curves are (hyper)elliptic. In this section, we discuss the spectral curve for
the $3\times3$ matrix linear ODE which appears in the theory of the
bi-orthogonal polynomials \cite{EMcL, BEH} with the cubic potentials
\cite{kapaev:bi3},
\begin{equation}\label{matrix_eqs_Psi}
\begin{split}
&\frac{\partial\Psi_n}{\partial\lambda}(\lambda)=
A_n(\lambda)\Psi_n(\lambda),\quad
\Psi_{n+1}(\lambda)=R_n(\lambda)\Psi_n(\lambda),
\\
&\frac{\partial\Psi_n}{\partial x}=U_n(\lambda)\Psi_n,\quad
\frac{\partial\Psi_n}{\partial y}=V_n(\lambda)\Psi_n,\quad
\frac{\partial\Psi_n}{\partial t}=W_n(\lambda)\Psi_n,
\end{split}
\end{equation}
where
$$
R_n(\lambda)=
\begin{pmatrix}
\frac{\lambda-a_{n,n}}{a_{n,n+1}}&
-\frac{a_{n,n-1}}{a_{n,n+1}}&
-\frac{a_{n,n-2}}{a_{n,n+1}}\\
1&0&0\\
0&1&0
\end{pmatrix},
$$
$$
A_n(\lambda)=-D_{n,n+2}R_{n+1}(\lambda)R_n(\lambda)
-D_{n,n+1}R_{n}(\lambda)
-D_{n,n}
-D_{n,n-1}R_{n-1}^{-1}(\lambda),
$$
$$
D_{n,m}=t\,\diag\bigl(
b_{n,m},b_{n-1,m-1},b_{n-2,m-2}\bigr).
$$
More details can be found in \cite{kapaev:bi3} where the
Riemann-Hilbert problem for $\Psi_n(\lambda)$ is formulated. The
fixed singularities of the relevant completely integrable system
correspond to the infinite values of the deformation parameters
$x,y,t$ as well as to $t=0$. In the case we are interesting here, the
formal monodromy exponents at $\lambda=\infty$, which is the only
singular point for the $\lambda$-equation in (\ref{matrix_eqs_Psi}),
are equal to $n,-n/2,-n/2$. The asymptotics of $\Psi_n(\lambda)$ as
$n\to\infty$ is of particular importance for the theory of coupled
random matrices. The scaling changes (\ref{scaling_change}) with
$\varkappa=-1/r_{\infty}=-1/3$ and Remark~\ref{scaling_freedom} imply
\begin{multline}\label{scaling_bi}
\lambda=\delta^{-1/3}\zeta,\quad
x=\delta^{-2/3}(x_0+\delta x_1),
\\
y=\delta^{-2/3}(y_0+\delta y_1),\quad
t=\delta^{-1/3}(t_0+\delta t_1),\quad
n=\delta^{-1},
\end{multline}
and yield the system (\ref{Psi_zeta}) with the spectral curve
\begin{multline}\label{spectral_curve_bi}
F(\zeta,\mu)=
\mu^3-t_0^3\zeta^3+\mu^2\zeta^2+x_0\mu^2+y_0t_0^2\zeta^2
-(t_0^3-1)\mu\zeta-
\\
-\mu D_1-\zeta D_2-D_3+{\cal O}(\delta)=0.
\end{multline}
Generically, this curve has 10 first order branch points and
therefore, via the Riemann-Hurwitz formula, has genus $g=3$. Because
the monodromy data for $\Psi_n(\lambda)$ are independent from
$n,x,y,t$, the curve (\ref{spectral_curve_bi}) for a generic solution
of (\ref{matrix_eqs_Psi}) satisfies (\ref{Boutroux}), see
Remark~\ref{Boutroux_importance}. By Theorem~\ref{Boutroux_D}, given
$x_0,y_0,t_0$, system (\ref{Boutroux}) uniquely determines the modular
parameters $D_j$, $j=1,2,3$.

\begin{figure}[hbt]
\begin{center}
\mbox{\psbox{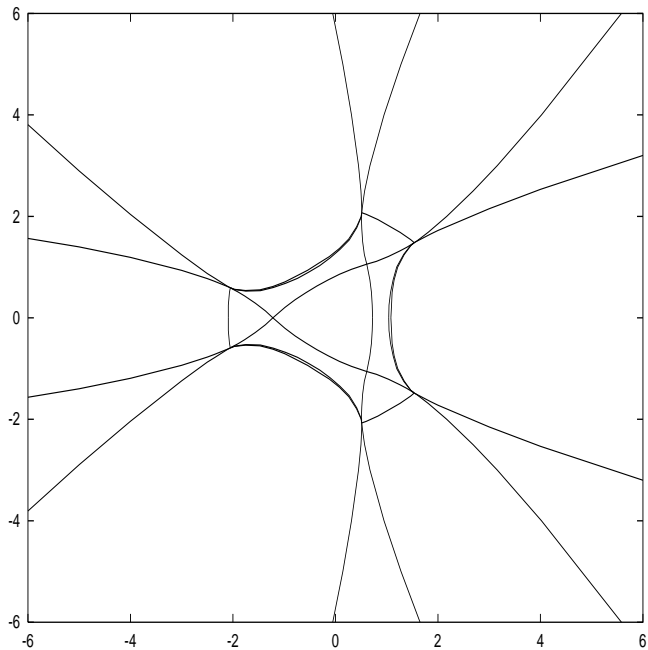}}
\end{center}
\caption{The projection of the integral sub-manifold
$y_0=\bar x_0$, $t_0=1$ for (\ref{Boutroux}) on the $x_0$-plane}
\end{figure}
The analysis of (\ref{Boutroux}) is significantly more involved then
the similar analysis of the elliptic curves associated to the
classical Painlev\'e equations, see \cite{kapaev:scls}. Here, we 
present the results in the numeric study of the integral sub-manifold
for (\ref{Boutroux}) parameterized by $x_0\in{\Bbb C}$ as 
$y_0=\bar x_0$ and $t_0=1$ based on the use of MATLAB~6.1 package.

The graph on the complex $x_0$-plane shown in Figure~1 separates the
regions with different topological properties of the relevant Stokes
graphs. Namely, some of the cycles $\ell_j$ existing in the
neighboring regions collapse at the points of their common
boundary. Our numeric study suggests that, at the points of the very
central triangular domain in Figure~1, the curve
(\ref{spectral_curve_bi}) subject to (\ref{Boutroux}) has genus $g=0$,
and the relevant Stokes graph is consistent with the Riemann-Hilbert
problem data of \cite{kapaev:bi3}. The latter observation implies
the applicability of (\ref{Boutroux}) to the asymptotic study of the
$\Psi$-function for the bi-orthogonal polynomials. In particular, the
typical configuration of the branch points implied by (\ref{Boutroux})
suggests that, for $x_0\neq0$, the asymptotics of $\Psi_n(\lambda)$
involves, in certain domains of the complex $\lambda$-plane, the
exponential, Airy and parabolic cylinder functions. For $x_0$ at the
boundary of the central triangular domain in Figure~1, the asymptotics
involves also the $\Psi$-function associated to the Painlev\'e first
transcendent. For $x_0=0$, besides exponential and Airy functions, the
asymptotic description requires also a third order special
function. For the values of $x_0$ beyond this triangular domain, the
relevant Stokes graphs seem not consistent with the Riemann-Hilbert
problem data of \cite{kapaev:bi3}. Therefore it is unlikely that, for
$x_0$ beyond the central triangular domain, the system
(\ref{Boutroux}) can be applied to the asymptotics of the
bi-orthogonal polynomials. The detailed description of the asymptotics
of the bi-orthogonal polynomials, however, is out of the scope of the 
present paper and will be published later elsewhere.

\bigskip
{\bf Acknowledgments.}
The work was supported in part by EPSRC and by RFBR (grant
N~02-01-00268). The author thanks A.~R.~Its for remarks, T.~Grava for
discussions and A.~S.~Fokas for support. The author is also grateful
to the staff of DAMTP, University of Cambridge, for hospitality during
his visit when this work was done.

\bibliographystyle{plain}
\ifx\undefined\bysame
\newcommand{\bysame}{\leavevmode\hbox to3em{\hrulefill}\,}
\fi

\end{document}

%% file: psbox.tex
\def\temp{1.31}
\let\tempp=\relax
\expandafter\ifx\csname psboxversion\endcsname\relax
\else
    \ifdim\temp cm>\psboxversion cm
    \else
      \let\temp=\psboxversion
      \let\tempp= 
    \fi
\fi
\tempp
\let\psboxversion=\temp
\catcode`\@=11
\def\execute#1{#1}
\def\psm@keother#1{\catcode`#112\relax}
\def\executeinspecs#1{%
\execute{\begingroup\let\do\psm@keother\dospecials\catcode`\^^M=9#1\endgroup}}
\def\psfortextures{
\def\PSspeci@l##1##2{%
\special{illustration ##1\space scaled ##2}%
}}
\def\psfordvitops{
\def\PSspeci@l##1##2{%
\special{dvitops: import ##1\space \the\drawingwd \the\drawinght}%
}}
\def\psfordvips{
\def\PSspeci@l##1##2{%
\d@my=0.1bp \d@mx=\drawingwd \divide\d@mx by\d@my%
\includegraphics{##1\space}%
}}
\def\psforoztex{
\def\PSspeci@l##1##2{%
\special{##1 \space
      ##2 1000 div dup scale
      \putsp@ce{\number-\psllx} \putsp@ce{\number-\pslly} translate
}%
}}
\def\putsp@ce#1{#1 }
\def\psfordvitps{
\def\psdimt@n@sp##1{\d@mx=##1\relax\edef\psn@sp{\number\d@mx}}
\def\PSspeci@l##1##2{%
\special{dvitps: Include0 "psfig.psr"}
\psdimt@n@sp{\drawingwd}
\special{dvitps: Literal "\psn@sp\space"}
\psdimt@n@sp{\drawinght}
\special{dvitps: Literal "\psn@sp\space"}
\psdimt@n@sp{\psllx bp}
\special{dvitps: Literal "\psn@sp\space"}
\psdimt@n@sp{\pslly bp}
\special{dvitps: Literal "\psn@sp\space"}
\psdimt@n@sp{\psurx bp}
\special{dvitps: Literal "\psn@sp\space"}
\psdimt@n@sp{\psury bp}
\special{dvitps: Literal "\psn@sp\space startTexFig\space"}
\special{dvitps: Include1 "##1"}
\special{dvitps: Literal "endTexFig\space"}
}}
\def\psforDVIALW{
\def\PSspeci@l##1##2{
\special{language "PS"
literal "##2 1000 div dup scale"
include "##1"}}}
\def\psonlyboxes{
\def\PSspeci@l##1##2{%
\at(0cm;0cm){\boxit{\vbox to\drawinght
  {\vss
  \hbox to\drawingwd{\at(0cm;0cm){\hbox{(##1)}}\hss}
  }}}
}%
}
\def\psloc@lerr#1{%
\let\savedPSspeci@l=\PSspeci@l%
\def\PSspeci@l##1##2{%
\at(0cm;0cm){\boxit{\vbox to\drawinght
  {\vss
  \hbox to\drawingwd{\at(0cm;0cm){\hbox{(##1) #1}}\hss}
  }}}
\let\PSspeci@l=\savedPSspeci@l
}%
}
\newread\pst@mpin
\newdimen\drawinght\newdimen\drawingwd
\newdimen\psxoffset\newdimen\psyoffset
\newbox\drawingBox
\newif\ifNotB@undingBox
\newhelp\PShelp{Proceed: you'll have a 5cm square blank box instead of
your graphics (Jean Orloff).}
\def\@mpty{}
\def\s@tsize#1 #2 #3 #4\@ndsize{
  \def\psllx{#1}\def\pslly{#2}%
  \def\psurx{#3}\def\psury{#4}
  \ifx\psurx\@mpty\NotB@undingBoxtrue
  \else
    \drawinght=#4bp\advance\drawinght by-#2bp
    \drawingwd=#3bp\advance\drawingwd by-#1bp
  \fi
  }
\def\sc@nline#1:#2\@ndline{\edef\p@rameter{#1}\edef\v@lue{#2}}
\def\g@bblefirstblank#1#2:{\ifx#1 \else#1\fi#2}
\def\psm@keother#1{\catcode`#112\relax}
\def\execute#1{#1}
{\catcode`\%=12
\xdef\B@undingBox{
}   
\def\ReadPSize#1{
 \edef\PSfilename{#1}
 \openin\pst@mpin=#1\relax
 \ifeof\pst@mpin \errhelp=\PShelp
   \errmessage{I haven't found your postscript file (\PSfilename)}
   \psloc@lerr{was not found}
   \s@tsize 0 0 142 142\@ndsize
   \closein\pst@mpin
 \else
   \immediate\write\psbj@inaux{#1,}
   \loop
     \executeinspecs{\catcode`\ =10\global\read\pst@mpin to\n@xtline}
     \ifeof\pst@mpin
       \errhelp=\PShelp
       \errmessage{(\PSfilename) is not an Encapsulated PostScript File:
           I could not find any \B@undingBox: line.}
       \edef\v@lue{0 0 142 142:}
       \psloc@lerr{is not an EPSFile}
       \NotB@undingBoxfalse
     \else
       \expandafter\sc@nline\n@xtline:\@ndline
       \ifx\p@rameter\B@undingBox\NotB@undingBoxfalse
         \edef\t@mp{%
           \expandafter\g@bblefirstblank\v@lue\space\space\space}
         \expandafter\s@tsize\t@mp\@ndsize
       \else\NotB@undingBoxtrue
       \fi
     \fi
   \ifNotB@undingBox\repeat
   \closein\pst@mpin
 \fi
\message{#1}
}
\newcount\xscale \newcount\yscale \newdimen\pscm\pscm=1cm
\newdimen\d@mx \newdimen\d@my
\let\ps@nnotation=\relax
\def\psboxto(#1;#2)#3{\vbox{
   \ReadPSize{#3}
   \divide\drawingwd by 1000
   \divide\drawinght by 1000
   \d@mx=#1
   \ifdim\d@mx=0pt\xscale=1000
         \else \xscale=\d@mx \divide \xscale by \drawingwd\fi
   \d@my=#2
   \ifdim\d@my=0pt\yscale=1000
         \else \yscale=\d@my \divide \yscale by \drawinght\fi
   \ifnum\yscale=1000
         \else\ifnum\xscale=1000\xscale=\yscale
                    \else\ifnum\yscale<\xscale\xscale=\yscale\fi
              \fi
   \fi
   \divide \psxoffset by 1000\multiply\psxoffset by \xscale
   \divide \psyoffset by 1000\multiply\psyoffset by \xscale
   \global\divide\pscm by 1000
   \global\multiply\pscm by\xscale
   \multiply\drawingwd by\xscale \multiply\drawinght by\xscale
   \ifdim\d@mx=0pt\d@mx=\drawingwd\fi
   \ifdim\d@my=0pt\d@my=\drawinght\fi
   \message{scaled \the\xscale}
 \hbox to\d@mx{\hss\vbox to\d@my{\vss
   \global\setbox\drawingBox=\hbox to 0pt{\kern\psxoffset\vbox to 0pt{
      \kern-\psyoffset
      \PSspeci@l{\PSfilename}{\the\xscale}
      \vss}\hss\ps@nnotation}
   \global\ht\drawingBox=\the\drawinght
   \global\wd\drawingBox=\the\drawingwd
   \baselineskip=0pt
   \copy\drawingBox
 \vss}\hss}
  \global\psxoffset=0pt
  \global\psyoffset=0pt
  \global\pscm=1cm
  \global\drawingwd=\drawingwd
  \global\drawinght=\drawinght
}}
\def\psboxscaled#1#2{\vbox{
  \ReadPSize{#2}
  \xscale=#1
  \message{scaled \the\xscale}
  \divide\drawingwd by 1000\multiply\drawingwd by\xscale
  \divide\drawinght by 1000\multiply\drawinght by\xscale
  \divide \psxoffset by 1000\multiply\psxoffset by \xscale
  \divide \psyoffset by 1000\multiply\psyoffset by \xscale
  \global\divide\pscm by 1000
  \global\multiply\pscm by\xscale
  \global\setbox\drawingBox=\hbox to 0pt{\kern\psxoffset\vbox to 0pt{
     \kern-\psyoffset
     \PSspeci@l{\PSfilename}{\the\xscale}
     \vss}\hss\ps@nnotation}
  \global\ht\drawingBox=\the\drawinght
  \global\wd\drawingBox=\the\drawingwd
  \baselineskip=0pt
  \copy\drawingBox
  \global\psxoffset=0pt
  \global\psyoffset=0pt
  \global\pscm=1cm
  \global\drawingwd=\drawingwd
  \global\drawinght=\drawinght
}}
\def\psbox#1{\psboxscaled{1000}{#1}}
\newif\ifn@teof\n@teoftrue
\newif\ifc@ntrolline
\newif\ifmatch
\newread\j@insplitin
\newwrite\j@insplitout
\newwrite\psbj@inaux
\immediate\openout\psbj@inaux=psbjoin.aux
\immediate\write\psbj@inaux{\string\joinfiles}
\immediate\write\psbj@inaux{\jobname,}
\immediate\let\oldinput=\input
\def\input#1 {
 \immediate\write\psbj@inaux{#1,}
 \oldinput #1 }
\def\empty{}
\def\setmatchif#1\contains#2{
  \def\match##1#2##2\endmatch{
    \def\tmp{##2}
    \ifx\empty\tmp
      \matchfalse
    \else
      \matchtrue
    \fi}
  \match#1#2\endmatch}
\def\warnopenout#1#2{
 \setmatchif{TrashMe,psbjoin.aux,psbjoin.all}\contains{#2}
 \ifmatch
 \else
   \immediate\openin\pst@mpin=#2
   \ifeof\pst@mpin
     \else
     \errhelp{If the content of this file is so precious to you, abort (ie
press x or e) and rename it before retrying.}
     \errmessage{I'm just about to replace your file named #2}
   \fi
   \immediate\closein\pst@mpin
 \fi
 \message{#2}
 \immediate\openout#1=#2}
{
\catcode`\%=12
\gdef\splitfile#1 {
 \immediate\openin\j@insplitin=#1
 \message{Splitting file #1 into:}
 \warnopenout\j@insplitout{TrashMe}
 \loop
   \ifeof
     \j@insplitin\immediate\closein\j@insplitin\n@teoffalse
   \else
     \n@teoftrue
     \executeinspecs{\global\read\j@insplitin to\spl@tinline\expandafter
       \ch@ckbeginnewfile\spl@tinline
     \ifc@ntrolline
     \else
       \toks0=\expandafter{\spl@tinline}
       \immediate\write\j@insplitout{\the\toks0}
     \fi
   \fi
 \ifn@teof\repeat
 \immediate\closeout\j@insplitout}
\gdef\ch@ckbeginnewfile#1
 \def\t@mp{#1}
 \ifx\empty\t@mp
   \def\t@mp{#3}
   \ifx\empty\t@mp
     \global\c@ntrollinefalse
   \else
     \immediate\closeout\j@insplitout
     \warnopenout\j@insplitout{#2}
     \global\c@ntrollinetrue
   \fi
 \else
   \global\c@ntrollinefalse
 \fi}
\gdef\joinfiles#1\into#2 {
 \message{Joining following files into}
 \warnopenout\j@insplitout{#2}
 \message{:}
 {
 \edef\w@##1{\immediate\write\j@insplitout{##1}}
 \w@{
 \w@{
 \w@{
 \w@{
 \w@{
 \w@{
 \w@{
 \w@{
 \w@{\string\input\space psbox.tex}
 \w@{\string\splitfile{\string\jobname}}
 }
 \tre@tfilelist#1, \endtre@t
 \immediate\closeout\j@insplitout}
\gdef\tre@tfilelist#1, #2\endtre@t{
 \def\t@mp{#1}
 \ifx\empty\t@mp
   \else
   \llj@in{#1}
   \tre@tfilelist#2, \endtre@t
 \fi}
\gdef\llj@in#1{
 \immediate\openin\j@insplitin=#1
 \ifeof\j@insplitin
   \errmessage{I couldn't find file #1.}
   \else
   \message{#1}
   \toks0={
   \immediate\write\j@insplitout{\the\toks0}
   \executeinspecs{\global\read\j@insplitin to\oldj@ininline}
   \loop
     \ifeof\j@insplitin\immediate\closein\j@insplitin\n@teoffalse
       \else\n@teoftrue
       \executeinspecs{\global\read\j@insplitin to\j@ininline}
       \toks0=\expandafter{\oldj@ininline}
       \let\oldj@ininline=\j@ininline
       \immediate\write\j@insplitout{\the\toks0}
     \fi
   \ifn@teof
   \repeat
   \immediate\closein\j@insplitin
 \fi}
}
\def\autojoin{
 \immediate\write\psbj@inaux{\string\into\space psbjoin.all}
 \immediate\closeout\psbj@inaux
 \input psbjoin.aux
}
\def\centinsert#1{\midinsert\line{\hss#1\hss}\endinsert}
\def\psannotate#1#2{\def\ps@nnotation{#2\global\let\ps@nnotation=\relax}#1}
\def\pscaption#1#2{\vbox{
   \setbox\drawingBox=#1
   \copy\drawingBox
   \vskip\baselineskip
   \vbox{\hsize=\wd\drawingBox\setbox0=\hbox{#2}
     \ifdim\wd0>\hsize
       \noindent\unhbox0\tolerance=5000
    \else\centerline{\box0}
    \fi
}}}
\def\psfig#1#2#3{\pscaption{\psannotate{#1}{#2}}{#3}}
\def\psfigurebox#1#2#3{\pscaption{\psannotate{\psbox{#1}}{#2}}{#3}}
\def\at(#1;#2)#3{\setbox0=\hbox{#3}\ht0=0pt\dp0=0pt
  \rlap{\kern#1\vbox to0pt{\kern-#2\box0\vss}}}
\newdimen\gridht \newdimen\gridwd
\def\gridfill(#1;#2){
  \setbox0=\hbox to 1\pscm
  {\vrule height1\pscm width.4pt\leaders\hrule\hfill}
  \gridht=#1
  \divide\gridht by \ht0
  \multiply\gridht by \ht0
  \gridwd=#2
  \divide\gridwd by \wd0
  \multiply\gridwd by \wd0
  \advance \gridwd by \wd0
  \vbox to \gridht{\leaders\hbox to\gridwd{\leaders\box0\hfill}\vfill}}
\def\fillinggrid{\at(0cm;0cm){\vbox{
  \gridfill(\drawinght;\drawingwd)}}}
\def\textleftof#1:{
  \setbox1=#1
  \setbox0=\vbox\bgroup
    \advance\hsize by -\wd1 \advance\hsize by -2em}
\def\textrightof#1:{
  \setbox0=#1
  \setbox1=\vbox\bgroup
    \advance\hsize by -\wd0 \advance\hsize by -2em}
\def\endtext{
  \egroup
  \hbox to \hsize{\valign{\vfil##\vfil\cr%
\box0\cr%
\noalign{\hss}\box1\cr}}}
\def\frameit#1#2#3{\hbox{\vrule width#1\vbox{
  \hrule height#1\vskip#2\hbox{\hskip#2\vbox{#3}\hskip#2}%
        \vskip#2\hrule height#1}\vrule width#1}}
\def\boxit#1{\frameit{0.4pt}{0pt}{#1}}
\catcode`\@=12 
 \psfordvips   